\documentstyle[preprint,aps]{revtex}

\begin{document}

\draft
\preprint{AZPH-TH/96-01}
\title{Phase Randomization and Doppler Peaks in the 
CMB Angular Power Spectrum}
\author{Li-Zhi Fang$^{1}$\cite{byline}, Zheng Huang$^{1}$\cite{dagger} and 
Xiang-Ping Wu$^{1,2}$\cite{ddagger}}
\address{\ $^1$
Department of Physics, University of Arizona, Tucson, AZ 85721, USA}
\address{\ $^2$ Beijing Astronomical Observatory, Beijing 100080, P.R.China}
\date{\today}
\maketitle
\begin{abstract}
Using the Boltzmann equation with a Langevin-like term describing the
stochastic force in a baryon-photon plasma, we investigate the influence 
of the incoherent electron-photon scattering on the subhorizon evolution of
the cosmic microwave radiation. The stochastic fluctuation caused by each
collision on average is found to be small. Nevertheless, it leads to a 
significant Brownian drifting of the phase in the acoustic oscillation,
and the coherent oscillations cannot be maintained during their dynamical
evolution. As a consequence, the proposed Doppler peaks probably do not exist.
\end{abstract}

\narrowtext

The anisotropies of the cosmic microwave background (CMB) provide a key to
the understanding of the origin of primordial fluctuations and the thermal 
history of the early universe. The spectrum of the CMB anisotropy on large 
angular scales has been found to be consistent with the inflationary 
scenario of the early universe\cite{1}. It is generally believed that many
cosmological parameters can be determined from the fine structures in the 
power spectrum of CMB anisotropy. Among them, the possible existence of
Doppler peaks -- the peaks in the CMB anisotropy spectrum on an angular 
scale of about one degree or less has attracted much attention\cite{2}.

It has been shown theoretically that the amplitude and the position of
these Doppler peaks are functions of the spatial curvatures, 
mass density, reionization time, cosmological constant etc. Thus, a 
precision measurement of the Doppler peaks may provide us an effective 
tool to determine various cosmic parameters. However, the  observed
amplitudes of the CMB anisotropy on degree-scale have not so far been
very conclusive in determining the existence of the Doppler peaks. Indeed,
some observations seems to exhibit high amplitudes as expected in a 
Doppler-peak scenario, while others show no peak amplitudes\cite{2}.
One may assert at this stage that the expected peaks have not yet been
clearly identified in current data, but would be determined by a new 
generations of the CMB anisotropy observations.

In this letter, we shall take a different approach to reexamine
the theoretical foundation of the prediction of the Doppler peaks. In
particular, we shall argue based on a stochastic Boltzmann equation that
the coherence of acoustic oscillations in the baryon-photon plasma will
be largely disturbed, and may even be totally erased if the stochastic
force of the incoherent scattering is included.
In the standard theory of the CMB evolution\cite{3}, the acoustic
oscillations in the baryon-photon plasma on subhorizon scales are treated
coherently, i.e. different modes are frozen at different phases of their
oscillation. The position of the Doppler peaks is then calculated by
the phase at the recombination. However, the inclusion of correlation
due to the stochastic term in the baryon-photon plasma will lead to a
phase randomization. The coherence of the oscillations could be maintained
if there is a mechanism for providing a negative entropy current to
prevent the decoherence due to the phase randomization (like a laser).
Unfortunately, no such mechanism exists in the epoch of recombination,
while the existence of a stochastic force term in the kinetic
equation is inevitable in a system with dissipation. We shall show that
the phase randomizations introduced by the incoherent electron-photon
scattering are, indeed, substantial, and as a result, the Doppler peaks
will be erased. Instead, one expects a large dispersion of 
the amplitudes in the CMB power
spectrum due to  different realizations on subhorizon scales,
reflecting the stochastic nature of the phases. We suggest that the
dispersed observations of the CMB angular power spectrum on
degree-scale, though very coarse at this stage, are in support of our view.

Let us use conventional notations in the theory of the CMB anisotropy\cite{3}.
Since the coherent oscillations of the Doppler peaks act only
on subhorizon scales, the choice of gauge is irrelevant for our
calculation. The evolution of the photon distribution function
$f(t, {\bf x}, {\bf p})$ is calculated by the Boltzmann equation
\begin{equation}
\frac{df}{dt} \equiv \frac{\partial f}{\partial t}
+ \frac{dx^i}{dt} \frac{\partial f}{\partial x^i}
+ \frac{dp_0}{dt} \frac{\partial f}{\partial p_0}
+ \frac{d\gamma^i}{dt} \frac{\partial f}{\partial \gamma^i}
=C[f].
\end{equation}
where $\gamma_i$ is the direction cosines of $p_i$ with respect to the
corresponding spatial coordinate. The left hand side of eq.(1) describes the
free-streaming, and the right hand side is the collision term given by
\begin{eqnarray}
C[f] & = & 
\int d{\bf q}d{\bf q'}d{\bf p'} W({\bf p}, {\bf q}, {\bf p'}, {\bf q'})
\nonumber \\
 &  & \times \left\{ g(t, {\bf x}, {\bf q'})f(t, {\bf x}, {\bf p'}) 
[1+f(t,{\bf x}, {\bf p})]  \right. \nonumber \\
 &  & \left. -g(t, {\bf x}, {\bf q})f(t, {\bf x}, {\bf p})
[1+f(t, {\bf x}, {\bf p'})] \right\} 
\end{eqnarray}
where $g$ is the electron distribution function, and the collision rate
$W$ is determined by the Compton scattering between electron and photon
from state $({\bf q'}, {\bf p'})$ to $({\bf q}, {\bf p})$.

It is well known that the Boltzmann equation in (1) is derived under the
assumption of a molecular chaos and is applicable if the
fluctuations caused by the incoherent collisions are negligible. 
These fluctuations give rise to a  stochastic force term
in hydrodynamics \cite{4}, governed by the fluctuation-dissipation theorem.
Similarly, these fluctuations can be taken into account by an additional
Langevin-like force term in the Boltzmann equation\cite{5}.
Eq.(1) should then be replaced by
\begin{equation}
\frac{df}{dt} = C[f] + r(t, {\bf x}, {\bf p})
\end{equation}
where the stochastic force, $r$, is characterized by its
correlations:
\begin{equation}
\langle r(t, {\bf x}, {\bf p}) \rangle = 0,
\end{equation}
\begin{eqnarray}
\langle r(t_1, {\bf x_1}, {\bf p_1})r(t_2, {\bf x_2}, {\bf p_2}) \rangle =
\frac{1}{2} N \delta(t_1 -t_2) \nonumber \\
 \times \delta({\bf x_1}-{\bf x_2})\int d{\bf q}d{\bf q'}d{\bf p}d{\bf p'}
W({\bf q}, {\bf p}, {\bf q'}, {\bf p'})
 \Delta({\bf p_1}) \nonumber \\
 \times  \Delta({\bf p_2})
 g(t_1, {\bf x_1}, {\bf q})f(t_1, {\bf x_1}, {\bf p})[1+
f(t_1, {\bf x_1}, {\bf p'})]
\end{eqnarray}
where $\langle ... \rangle$ is an average over the stochastic effect
(i.e.\ over different realizations),
$N$ is the
total number of photons, and the function $\Delta({\bf P})$ is defined by
\begin{eqnarray}
\Delta({\bf P}) & =  & \delta({\bf P}-{\bf q}) + \delta({\bf P}-{\bf p})
\nonumber \\
  & & - \delta({\bf P}-{\bf q'}) - \delta({\bf P} - {\bf p'}).
\end{eqnarray}
It is important to note that eqs.(3)-(6) are applicable not only in 
a linear approximation but also in the nonlinear region\cite{5,6}.

In principle, to study the fluctuations in the  baryon-photon plasma, we
should also consider the stochastic terms in the equation of baryonic 
matter. However, since the linear fluctuations given by independent 
Gaussian random ``forces" are additive, the stochastic terms in the baryonic
equation will increase the effects of fluctuations considered in this 
paper. To illustrate the main effect of electron-photon scattering in 
the presence of the Doppler peaks, we shall only consider the
contribution from stochastic term $r$ in photon's equation. In this case,
the electron distribution $g(t, {\bf x}, {\bf q})$ can be treated
as an external source. For a Thomson scattering, photons do not exchange
energies with electrons. The ${\bf q}$-distribution of electrons do not
involved in the evolution, and we need only the number density distribution
of electrons $n_e=\int d{\bf q}g(t, {\bf x}, {\bf q})$. In this case,
the distribution of the photon energy $p$ is also unchanged,
the perturbation of the CMB can be described by the anisotropy of the
brightness temperature
\begin{equation}
\Theta(t, {\bf x}, {\bf \gamma})
=\frac {1}{4\pi^2\rho_{\gamma}} \int f p^3dp-\frac {1}{4},
\end{equation}
where $\rho_{\gamma} = (\pi^2/15)T^4$ is the mean energy density of 
photons. In terms of $\Theta$, the photon distribution function can 
be approximately expressed as
\begin{equation}
f(t, {\bf x}, {\bf p}) = f_T(p)[1+4\Theta(t, {\bf x}, {\bf \gamma})]
\end{equation}
where $f_T= 1/[\exp(p/T)-1]$. 

 From eqs.(1) and (8), one finds that $\Theta$ should satisfy
\begin{eqnarray}
 \frac{\partial}{\partial \eta} (\Theta + \Phi)
 +  \gamma^i \frac{\partial }{\partial x^i}(\Theta + \Psi)
+ \dot{\gamma^i}\frac{\partial}{\partial \gamma^i} \Theta = \nonumber \\
   \tau (\Theta_0 - \Theta + \gamma_i v_b^i + \frac{1}{10}{\bf Q} \Theta )
   + R
\end{eqnarray}
where $\Phi$ and $\Psi$ are the Newtonian potential and the space curvature
perturbation, respectively. $\eta=\int (1+z)dt$ is the conformal time, $z$
the redshift, $v_b$ the baryonic (fluid) velocity, and $\tau = n_e\sigma_T$ 
the optical depth, $\sigma_T$ being the Thomson cross section. ${\bf Q}$ is
the projection operator of quadrupole, defined as
${\bf Q(\gamma, \gamma')} = \sum_{m=-2}^{2}Y^*_{2m}({\gamma})
(1/4\pi)\int d\Omega'Y_{2m}({\bf \gamma'})$.
Obviously {\bf Q}{\bf Q}={\bf Q}. The isotropic component
$\Theta_0(\eta, {\bf x})$ is
given by $\Theta_0= {\bf O} \Theta $, where
${\bf O}\equiv (1/4\pi)\int d{\Omega}$
is the monopole projection operator, and {\bf O}{\bf O}={\bf O}.
The term $\Theta_0$ appearing on the  right-hand side of eq.(9) 
indicates that  without an external driving force
(such as $v_b$) the isotropic state is the ``equilibrium'' state in the
kinetic evolution of the Thomson scattering.

The stochastic term in eq.(9) is $R=(1/4\pi^2\rho_r) \int r p^3dp$, and its
correlation function is given by
\begin{eqnarray}
 \langle R(\eta_1, {\bf x_1}, {\bf \gamma_1})
 R(\eta_2, {\bf x_2}, {\bf \gamma_2}) \rangle = \nonumber \\
 \frac{3}{8\pi^4} \frac{N}{16\pi^4\rho_r^2} \tau \delta(\eta_1 -\eta_2) 
  \delta({\bf x_1}-{\bf x_2}) \nonumber \\
 \times [\frac{4}{3} 4\pi \delta ({\bf \gamma_1} - {\bf \gamma_2})
 - \frac{1}{2}(1+\cos^2 \beta)]       \nonumber \\
 \times \int p^4dp [f(\eta_1, {\bf x_1}, p, {\bf \gamma_1}) \nonumber \\
 + 2f(\eta_1, {\bf x_1}, p, {\bf \gamma_1})
 f(\eta_2, {\bf x_2}, p, {\bf \gamma_2})
 + f(\eta_2, {\bf x_2}, p, {\bf \gamma_2})] 
\end{eqnarray}
where $\beta$ is the angle between ${\bf \gamma_1}$ and ${\bf \gamma_2}$.
Because eq.(9) is linear in $\Theta$, it reduces to 
eq.(1) by taking an average over the stochastic effect. Thus,
the calculation based on eq.(1) is actually  only for
 $\langle \Theta \rangle$ but not for $\Theta$, i.e. the fluctuations due
to the stochastic terms are entirely overlooked in eq.(1). In the
incoherent processes, the linear fluctuations,
$\delta \Theta = \Theta - \langle \Theta \rangle$, 
caused by stochastic force $R$ may not in general play a very important 
role. However, these fluctuations lead to ``forgetting history''. Thus, the
existence and maintenance of coherence should be seriously reconsidered.

Let us calculate the linear fluctuations, 
$\delta \Theta(\eta, {\bf x}, \gamma)$,
at a given time $\eta$ around a solution to eq.(1). It is governed by 
\begin{eqnarray}
  \left \{ \frac{\partial }{\partial \eta} +
\gamma^i \frac{\partial }{\partial x^i}
+ \dot{\gamma^i}\frac{\partial }{\partial \gamma^i}  
\right. &  & \nonumber \\
  \left. +  \tau [1 + \frac{1}{10} {\bf Q}] \right \}
(\delta \Theta -\delta \Theta_0) & = & R'
\end{eqnarray}
where $R'=(1- {\bf O})R$. The projection operator arises from
the isotropic term $\Theta_0$ in eq.(9). The solution to eq.(12) can
be generally expressed as
\begin{eqnarray}
\delta \Theta(\eta, {\bf x}, {\bf \gamma}) - 
\delta \Theta_0(\eta, {\bf x}, {\bf \gamma})=
\int_0^{\infty} d\lambda e^{-\lambda \tau[1+ (1/10){\bf Q}]} \nonumber \\
\times R'(\eta - \lambda, x^i - \int_0^{\lambda} d\lambda ' \gamma^i,
\gamma^i - \int_0^{\lambda} d\lambda ' \dot{\gamma^i})
\end{eqnarray}
Obviously,  both sides of eq.(13) become zero upon an integration
$(1/4\pi)\int d\Omega$ because ${\bf O}(1 - {\bf O}) =0$.

The correlation functions of fluctuations, such as
$\langle \delta\Theta(\eta_1, {\bf x_1}, {\bf \gamma_1})
\delta\Theta(\eta_2, {\bf x_2}, {\bf \gamma_2}) \rangle$, can be calculated
from eqs.(10) and (12). For a flat universe, the light path is a
straight line in comoving coordinates
${\bf x}$, and thus $\dot{\gamma^i} = 0$. The fluctuation for the
mode $(l, {\bf k})$ with $l>0$ is then
\begin{eqnarray*}
\langle \delta\Theta_l(\eta, {\bf k})\delta\Theta_l^*(\eta, {\bf k})\rangle
=   \nonumber \\
\frac{1}{V^2} \int d{\bf x_1} d{\bf x_2}
\exp\{-i{\bf k}\cdot({\bf x_1}-{\bf x_2})\}  \nonumber \\
\frac{(2l+1)^2}{64\pi^2}
\int d\Omega_1 d\Omega_2 P_l({\bf k_0} \cdot {\bf \gamma_1})
P_l({\bf k_0}\cdot {\bf \gamma_2})  \nonumber \\
 \int d\lambda_1 d\lambda_2
\exp\{ - \tau\lambda_1[1+ {\bf Q}_1] - \tau\lambda_2[1+ {\bf Q}_2]\} \nonumber \\
\langle
R'(\eta-\lambda_1,{\bf x_1} -\lambda_1{\bf \gamma_1},{\bf \gamma_1})
R'(\eta-\lambda_2,{\bf x_2}\ - \lambda_2 {\gamma_2},{\bf \gamma_2})
\rangle
\end{eqnarray*}
where ${\bf k_0}= {\bf k}/|{\bf k}|$, and $V$ the volume being considered.
Since $(\tau/ k) \gg 1$, the above equation can be simplified by using eqs.(8)
and (10) as
\begin{eqnarray} 
\langle \delta \Theta_l(\eta, {\bf k})
\delta \Theta_l^*(\eta, {\bf k}) \rangle =  \frac{0.531 \times 675}{\pi^{14}}
\nonumber \\
\times \frac{(2l+1)^2}{64\pi^2}
\int d\Omega_1 d\Omega_2 P_l({\bf k_0} \cdot {\bf \gamma_1})
P_l({\bf k_0}\cdot {\bf \gamma_2}) \nonumber \\
 \int \tau d\lambda e^{ -2\tau \lambda} (1-{\bf O_1})(1- {\bf O_2}) \nonumber \\
 \times [1+(e^{\tau\lambda/10}-1){\bf Q_1}]
 [1+(e^{\tau\lambda/10}-1){\bf Q_2}]  \nonumber \\
 \times [\frac{4}{3} 4\pi \delta({\bf \gamma_1} - {\bf \gamma_2})
-\frac{1}{2}(1+\cos^2\beta)]  \nonumber \\
\times \frac{1}{V}\int d{\bf x}\langle \Theta(\eta-\lambda, {\bf x}, {\bf \gamma_1})
\Theta^*(\eta-\lambda, {\bf x}, {\bf \gamma_2})\rangle
\end{eqnarray}
where we used $N=(2\zeta(3)/\pi^2) VT^3$, 
$\int p^4 dp f^2_T= 2[\zeta(2)-\zeta(3)]T^5= 0.844 T^5$, and $e^{a{\bf Q}}=
[1+(e^a -1){\bf Q}]$.
The monopole projection factors $(1-{\bf O_1})$
and $(1-{\bf O_2})$ can be replaced by a unity operator when $l>0$.
The two quadrupole projection terms ${\bf Q_1}$ and ${\bf Q_2}$ can
be neglected  because the term $e^{-2\lambda \tau}$ picks up the main
contribution in the integral from the region 
$\lambda \tau < 1$. Using the following mode decomposition
\begin{eqnarray}
\frac{1}{V}\int d{\bf x}\langle \Theta(\eta, {\bf x} , {\bf \gamma_1})
\Theta^*(\eta, {\bf x}, {\bf \gamma_2}) \rangle = \nonumber \\
\frac{V}{2\pi^2}\int_0^{\infty} k^2dk \sum_{l'}
\frac{1}{2l'+ 1} \langle |\Theta_{l'}(\eta, k)| ^2\rangle
P_{l'}({\bf \gamma_1}\cdot {\bf \gamma_2})
\end{eqnarray}
we have finally
\begin{eqnarray}
\langle |\delta \Theta_l(\eta, {\bf k})|^2 \rangle
 =  3.91\times 10^{-5} \nonumber \\ 
 \left\{ \frac{2l+1}{3}\frac{V}{2\pi^2}\int dkk^2
 \sum_{l'}\frac{1}{2l'+1}\overline{\langle |\Theta_{l'}(\eta, k)|^2 \rangle}
  \right. \nonumber \\
 - \frac{2}{3} \sum_{\nu=0, \pm 2}[\delta_{\nu,0}
 + |(l+\nu,0, 2, 0|l+\nu,0)|^2]  \nonumber \\
 \left. \times \frac{V}{2\pi^2}
 \int k^2dk\frac{1}{2l+1}
\overline{\langle |\Theta_l(\eta, k)| ^2\rangle } \right\}
\end{eqnarray}
where the overline above $\langle |\Theta_l(\eta, k)| \rangle ^2$
denotes an average over the region from $\eta - (1/\tau)$ to $\eta$,
and $(l+\nu,0,2,0|l+\nu,0)$ are the Clebsch-Gordon coefficients.
For large $l$, the second (negative) term in the bracket eq.(16) is
completely negligible. The first term is independent of $l$ and $k$. This is
expected since the correlation function of the corresponding stochastic force
is isotropic in $\gamma$-space and uniform in $k$-space. 

Using the expression for a temperature perturbation
\begin{eqnarray}
\left(\frac{\Delta T}{T}\right)^2=
\frac{V}{2\pi^2}\int dk k^2
\sum_{l}\frac{1}{2l+1}\overline{\langle |\Theta_{l}(\eta, k)|^2\rangle},
\end{eqnarray}
one has
\begin{equation}
\frac{\langle |\delta \Theta_l|^2 \rangle}{2l+1} \simeq 1.30 \times 10^{-5}
|\Delta T/T|^2.
\end{equation}
The $l$ summation in eq.(16) should run up to $l_{max}$ where the
electron-photon collision is frequent enough, i.e. to the scale about
the photon  mean free path $1/\tau$. Therefore, $\Delta T/T$ in
eqs.(16) and (17) should not be confused with $(\Delta T/T)_{obs}$ given
by observations with low resolutions, i.e. their window functions are
on scales much larger than $l_{max}$. For the observation of CMR
anisotropy with a resolution $l$, the fluctuation is
\begin{eqnarray}
\lefteqn{ \delta \left(\frac{\Delta T}{T} \right) =
\left( \sum_{l'=0}^{l}\frac {\langle |\delta\Theta_{l'}|^2
\rangle}{2l'+1} \right)^{1/2} }
\nonumber \\
 & & \simeq 3.61 \times 10^{-3} \sqrt{l} \frac{\Delta T}{T}.
\end{eqnarray}
Therefore, the fluctuations caused by the stochastic force $R$  
are generally small, except for the case of a very high resolution
observation. 

However, the average of the fluctuations in eqs.(15) and (18) is essentially 
over one collision time $1/\tau$. In relation to coherent processes, we
should study the cumulative effect of the fluctuations over the entire
period during which the coherence is to be maintained\cite{7}. For
the Doppler peaks, the period is from the time 
when the perturbation enters horizon
to the time of recombination, 
i.e.\ the duration of the subhorizon evolution before
recombination.
The cumulative effect can be easily described by the phase fluctuations of
the oscillation. Use the expression
$\Theta_l(\eta, k)=|\Theta_l|e^{i\phi_l}$, where
$|\Theta_l(\eta, k)|$ and $\phi_l(\eta, k)$ are the amplitude and phase,
respectively. The equations of $\Theta_l(\eta, k)$ can then
be derived from eq.(9) as
\begin{eqnarray}
\dot {\Theta_0} = -\frac{k}{3} \Theta_1 - \dot \Phi + R_0  \\
\dot {\Theta_1} = k \left(\Theta_0 + \Psi - \frac{2}{5}\Theta_2 \right) -
\tau (\Theta_1 - V_b) + R_1 \\
\dot {\Theta_l} = k \left(\frac{l}{2l-1}\Theta_{l-1}-
\frac{l+1}{2l+3}\Theta_{l+1}\right)  \nonumber \\
  - \tau \Theta_l + R_l   \hspace{3mm} (l>2)
\end{eqnarray}
where
\begin{equation}
R_l= \frac{1}{V}\frac{2l+1}{8\pi}
\int d\Omega P_l({\bf k_0}\cdot{\bf \gamma})
\int d{\bf x} e^{-i{\bf k}\cdot{\bf x}} R
\end{equation}
If the hierarchy of (21)-(23) is cut off at 2$l$-th order, they correspond to
the equations for a system consisting of $l$ coupled oscillators. Without the
stochastic terms $R_l$, the oscillations are coherent and the phases of
the oscillations are completely fixed by  initial conditions.
The terms $R_l$ lead to a phase randomization.

We shall first calculate the phase fluctuations raised by $R_0$. As the
coupling between the photon and the baryon
is tight and the peculiar gravitational potential is weakly time-dependent
in matter-dominated regime\cite{3}, the evolution of $\Theta_0$ is dominated
by the phase evolution $\phi_0(\eta)$, 
which is approximately described by a WKB-like
equation as
\begin{equation}
\frac{d\phi_0}{d\eta} \simeq kc_s - 
\frac{1}{2\langle\Theta_0 \rangle}\frac{1}{kc_s}\frac {dR_0}{d\eta} \ ,
\end{equation}
where $c_s$ denotes the sound speed of the baryon-photon plasma, which is
$\sim 1/\sqrt{3}$ before recombination.
The stochasticly averaged solution of eq.(23) is
$\phi_0(\eta)=\int_{\eta_{en}}^{\eta}kc_sd\eta' + \phi_0(\eta_{en})$,
where $\eta_{en}$ denotes the time when the mode $k$ enters the horizon,
and $\phi_0(\eta_{en})$ is the initial phase. The position of the
Doppler peaks for adiabatic perturbations is approximately determined by
a phase relation $\phi_0(\eta_{re}) = n\pi$, where $n$ is integer.
Eq.(23) shows also that the phase fluctuation due to $R_0$
in the period from $\eta_{en}$ to $\eta_{re}$ is
\begin{equation}
\delta\phi_0 \simeq \int_{\eta_{en}}^{\eta_{re}} d\eta'
\frac{1}{2\langle \Theta_0 \rangle } \frac{1}{kc_s}
\frac {dR_0(\eta')}{d\eta'} \ .
\end{equation}

Generally, the term $R_l$ leads to a phase fluctuation at least of
\begin{equation}
\delta\phi_l \simeq \int_{\eta_{en}}^{\eta_{re}} d\eta'
\frac{1}{2\langle \Theta_l \rangle } \frac{1}{kc_s}
\frac {dR_l(\eta')}{d\eta'} \ .
\end{equation}
Considering that the derivative $(1/kc_s)d/d\eta$ in eq.(25)
contributes a factor
of order 1, and using eqs.(10), (15) and (17), one obtains the mean phase
fluctuation  as
\begin{eqnarray}
\langle (\delta \phi_l)^2 \rangle \simeq
\int_{\eta_{en}}^{\eta_{re}}
\frac{1}{4\langle|\Theta_l|^2 \rangle}
\langle |\delta\Theta_l(\eta, k)|^2 \rangle \tau d\eta \nonumber \\
\simeq 0.33 \times 10^{-5} (2l+1) \int_{\eta_{en}}^{\eta_{re}}
\frac{1}{\langle|\Theta_l|^2 \rangle}
\left(\frac{\Delta T}{T} \right)^2 \tau d\eta
\end{eqnarray}
The factor
$(\Delta T/T)^2/\langle |\delta\Theta_l(\eta, k)|^2 \rangle$ is no
less than 1, and therefore, the RMS of the phase fluctuation can be
estimated as
\begin{equation}
\sqrt{\langle (\delta \phi_l)^2 \rangle} \geq
1.8\times 10^{-3} \sqrt{(2l+1)N_c}
\end{equation}
where $N_c \equiv \tau(\eta_{re}-\eta_{en})$ is the mean number of
 collisions in
the entire period of subhorizon evolutions of mode $k$ before the
recombination.
Therefore, the behavior of the stochastic fluctuation of the phase for the
$l$-oscillation is just like a Brownian drifting: the number of collisions
corresponds to the
number of steps in the random walk, and the mean shift per  step is about
$1.8\times 10^{-3}\sqrt{2l+1}$. For Doppler peaks, we have $l \geq 100$,
and $N \geq \tau(\eta_{re}-\eta_{en}) \geq 10^3$. Therefore, we conclude 
that the RMS of the Brownian phase drifting due to Thomson scattering 
is order 1, significant enough to disturb the coherence.

In addition to the Thomson scattering, there are other stochastic forces
in the baryon-photon plasma, such as non-Thomson terms
of the Compton scattering, the stochastic terms in the hydrodynamical 
equation of baryons. Fluctuations from different sources mostly are 
additive. Our estimation on the  phases drifting is thus a conservative
one. The proposed Doppler peaks in the CMB angular power spectrum 
probably do not exist. Different causal areas at the recombination can 
be considered as independent realizations of the stochastic force. The
dispersion of the currently observed CMB anisotropies on angular scale of 
one degree is consistent with the scenario of the Brownian drift of the 
phases.

This work (Z.H.) was supported in part through the U.S. Department 
of Energy under Contract Nos. DE-FG03-93ER40792 and DE-FG02-85ER40213.
X.P.W. is supported by a World Laboratory Fellowship.

\end{document}